\documentclass[a4paper]{article}

\usepackage[english]{babel}
\usepackage[T1]{fontenc}

\usepackage[a4paper,top=3cm,bottom=2cm,left=3cm,right=3cm,marginparwidth=1.75cm]{geometry}

\usepackage{amsmath}
\usepackage{graphicx}

\title{Screening for cancer using a learning Internet advertising system}
\author{Elad Yom-Tov\\Microsoft Research Israel\\eladyt@microsoft.com}
\date{}

\begin{document}
\maketitle

\begin{abstract}

Studies have shown that the traces people leave when browsing the internet may indicate the onset of diseases such as cancer. Here we show that the adaptive engines of advertising systems in conjunction with clinically verified questionnaires can be used to identify people who are suspected of having one of three types of solid tumor cancers.

In the first study, 308 people were recruited through ads shown on the Bing search engine to complete a clinically verified risk questionnaire. A classifier trained to predict questionnaire response using only past queries on Bing reached an Area Under the Curve of 0.64 for all three cancer types, verifying that past searches could be used to identify people with suspected cancer.

The second study was conducted using the Google ads system in the same configuration as in the first study. However, in this study the ads system was set to automatically learn to identify people with suspected cancer. A total of 70,586 people were shown the ads, and 6,484 clicked and were referred to complete the clinical questionnaires. People from countries with higher Internet access and lower life expectancy tended to click more on the ads. Over time the advertisement system learned to identify people who were likely to have symptoms consistent with suspected cancer, such that the percentage of people completing the questionnaires and found to have suspected cancer reached approximately 11\% at the end of the experiment.  

These results demonstrate the utility of using search engine queries to screen for possible cancer and the application of modern advertising systems to help identify people who are likely suffering from serious medical conditions. This is especially true in countries where medical services are less developed.

\end{abstract}

\section{Introduction}

Data generated while people browse the Internet, especially when using Internet search engines, has been shown to reflect the experiences of people in the physical world \cite{yomtov2016crowdsourced}. Indeed, the vast majority of Internet users refer to search engines when they have a medical concern \cite{pew2013}. For this reason, search queries have been used to track infectious diseases such as influenza \cite{polgreen2008,lampos2015}, answer questions on the relationship between diet and chronic pain \cite{giat2018}, and to identify precursors to disease \cite{yomtov2015}. 

Recently, search engine queries were shown to be a useful signal for evaluating whether people are likely to be suffering from different cancer types, including ovarian \cite{soldaini2017}, cervical \cite{soldaini2017}, pancreatic \cite{paparrizos2016}, and lung \cite{white2017}. The underlying assumption in all these studies is that these cancers manifest themselves in externally recognizable symptoms that are either unfamiliar or relatively benign, meaning that people do not immediately turn to professional medical consultation.  
 
A basic shortcoming in studying the connection between web searches and disease onset is that web searches are anonymous. This means that linking actual medical information, such as exact disease diagnosis and date, is limited to indirect inference, such as queries of Self Identified Users (SIUs).  SIUs are people who, in their queries, identify themselves as having a condition of interest, e.g., "I have breast cancer" (See \cite{white2017,paparrizos2016}). Unfortunately, people who self-identify are few and are drawn from an unrepresentative population \cite{soldaini2017,yomtov2018demog}. Others \cite{soldaini2017} used both SIUs and information on geographic variability in disease incidence to infer which users are suffering from specific cancers, based on the queries they made. This method provided a larger (and more diverse) cohort than is possible with SIUs, but is still lacking in that no clinical information about the users is known. This makes it impossible to have a definite clinical indicator of disease.

An approach to obtaining clinical indicators of disease is through questionnaires. de Choudhury et al. \cite{choudhury2013} used questionnaires to assess the level of depression of crowdsourced workers, as a basis for using their social media posts to distinguish depressed from non-depressed individuals. This study focuses on cancer and correlates the score of a clinically validated questionnaire to medical symptom searches. Since the incidence of cancer is significantly lower than that of depression (creating a recruitment challenge) and people are more likely to ask for medical symptoms, especially those of a personal nature, on search engines rather than social media \cite{pelleg2012}, we used a targeted advertising campaign and search engine logs as our primary data source. 

Operating in conjunction with all major search engines is a highly developed advertising system. In this work we show, for the first time, that the ads serving platform that accompanies search engines can be used to target population at risk. This is done by utilizing attributes of the ads systems to automatically learn to screen for three types of cancer: Breast, colon, and lung. This could provide a major public health benefit, especially in countries with under-developed access to healthcare. 

Advertising systems show ads to people when they use a search engine to query for terms defined by the advertisers. These ads  display a short text (and sometimes an image) and provide a link to the advertiser's website. Advertisers commonly pay whenever a user clicks on the ad, and therefore the ads platform is optimized to show a certain ad only when it is likely to be clicked. More recently, advertising systems have begun allowing advertisers to signal the system when a user purchases the advertised product. This indication, known as a {\bf conversion}, allows the search engine to use past searches and other parameters to identify people who are likely to purchase the product, not just to click on an ad. Over time, a system can learn to identify such people from the feedback provided by the advertiser. Ipeirotis and Gabrilovich \cite{Ipeirotis2014} used this mechanism to find people who can correctly answer questions on topics of interest, by providing a conversion signal when people answered several test questions correctly. 

Thus, here we propose to utilize this mechanism as follows: Users asking if they are likely to have specific cancers will be referred through an ad to a clinically validated questionnaire which calculates the likelihood of the user having a specific type of cancer. This score will be provided to users, and if the score is such that the person is likely to have cancer, will also be provided as a conversion signal to the advertisement system. Thus, we utilize the questionnaire score as a conversion signal. Our hypothesis is that over time the system will learn to identify more people who will score high on the questionnaires, indicating that more people at risk of cancer are identified. %Consider adding a figure

Thus, our contributions here are threefold: First, we use clinically verified questionnaires to calculate the likelihood of a user having a suspected cancer type in lieu of queries which indicate a cancer diagnosis (self identified users). Second, we correlate individual questionnaire scores of people with their past search engine queries to show that suspected cancer could have been predicted based on these queries. Finally, we use the learning capabilities of advertising systems to find more people who are likely to have suspected cancer. Our results demonstrate that the proposed methods can assist in finding people with suspected cancer in an accurate and economic manner.

\section{Methods}

\subsection{Overview}

We focused on three types of cancer: Lung, breast, and colon. These three were chosen for their relatively high incidence and because the symptoms of these cancers (as described in the relevant questionnaires) were assessed to be understood by laypersons. 

Users were recruited through ads shown when they searched for information on diagnosis of these three specific cancers. People who clicked on these ads were referred to a specially designed website where clinically validated questionnaires on whether they should see a specialist oncologist were administered to them. The scores were provided to users. Users were requested to provide their data for the experiment. In the first study the scores were correlated with past searches of users who consented. In the second study, a conversion indicator was given to the advertising system for users with high scores.  

The first study was conducted using the Bing ads system, and required privileged access to the search system data to obtain past user queries. The second study was conducted using the Google ads system, with no such privileged access to past queries. The latter was done to demonstrate that public health organizations with no privileged access could also utilize these systems. In both studies the campaign budget was set to US\$15 per day, which meant that not all people who issued the relevant queries could be shown the ads. This was done so as to allow the ads system in the second study to select relevant participants from all users issuing relevant queries.

This study was approved by the Microsoft Institutional Review Board (IRB9672).

\subsection{Recruitment}

Recruitment was similar in both studies: Users were recruited through ads displayed using the respective ads system. Recruitment ads were shown when people searched for "symptoms of <cancer type>", "signs of <cancer type>", "<cancer type> diagnosis", "<cancer type> quiz", or "<cancer type> questionnaire".   

The ads contained one of the following three titles: "<cancer type> - Do you have it?", "<cancer type> - Think you have it?" or "<cancer type> - Worried you have it?". The text of the ads was "Click here to check if you should see a doctor" (or physician). All ads were shown with equal probability.

\subsection{Questionnaires}

People who clicked on these ads were referred to a specially designed website. In this website they were shown a questionnaire developed by the UK National Institute for Health and Care Excellence (NICE). The Suspected Cancer Recognition and Referral questionnaires were designed to assist general practitioners to decide if a patient should see a specialist oncologist. After answering the questions on the questionnaire, physicians are advised whether or not to refer people for an appointment with the specialist within two weeks. We refer to the the output of the questionnaire as a {\bf suspected cancer score} (SCS). People with a high SCS are advised to consult an oncologist within two weeks. 

In this work, users who responded to the questionnaire on the website and received a high SCS were advised to consult with a doctor immediately (in the clinical use of the questionnaire patients with high score are referred to an oncologist within 2 weeks).  If the SCS was low, users are advised that their symptoms were not commonly associated with cancer but that they should see a medical doctor if the symptoms were persistent or worrying. 

Users were asked for their consent to participate in the study {\bf both} at the beginning of the questionnaire and after the results of the questionnaire were provided to them. Only people who consented in both times were included in the study. 

\subsection{Study 1: Suspected cancer scores are correlated with past search engine queries}

People were recruited through ads, and asked to complete questionnaires. The ads were shown between December 29th, 2017 and March 31st, 2018 to people in the USA. We extracted all queries made on Bing by users who completed a questionnaire, consented to contribute their data to the study, and who were logged in to Bing at the time of ad display. The queries were extracted from 3 months before the questionnaire completion and until that date.  Queries of each person were represented by:
\begin{enumerate}
\item The number of times a medical symptom was mentioned in the queries. Symptoms were comprised of a list of 195 symptoms and their layperson description as developed in \cite{yomtov2013}.
\item The words and word pairs (excluding stopwords) in the queries, if these phrases appeared in use by at least 5\% of people in the sample. 
\end{enumerate}

The SCS was predicted from query data of participants for whom at least 14 days of query data were available.  Patterns (independent attributes) for prediction included the query terms, as described above, as well as age and gender of the user. We used a random forest \cite{breiman2001}, and evaluated the performance of the model using leave-one-out estimation \cite{duda2012}. 

\subsection{Study 2: Advertising systems can learn to identify people with suspected cancer}

Ads were shown on the Google ads system between May 16th and June 12th, 2018. Once each questionnaire was completed, and if users consented to participating in the study, a conversion signal was fed to the Google advertising system for those users who's SCS was high. We report the conversion rate over time, that is, the percentage of people who saw the ads, clicked on them, and were found to have a high SCS. If the system can learn to identify people with high SCS, it is expected that this rate will rise over time.

\section{Results}

\subsection{Study 1: Suspected cancer scores are correlated with past search engine queries}

The experiment was run between December 7th, 2017 and April 13th, 2018. During this time recruitment ads were shown 159,170 times and clicked 2,899 times. Clickthrough rates for different conditions were similar, ranging from 1.2\% (breast cancer) to 4.8\% (colon cancer). Females and males were similarly likely to click on the ads for colon and lung cancer, but females were 2.0 times more likely to click on ads for breast cancer.

\begin{figure}
  \includegraphics[width=\linewidth]{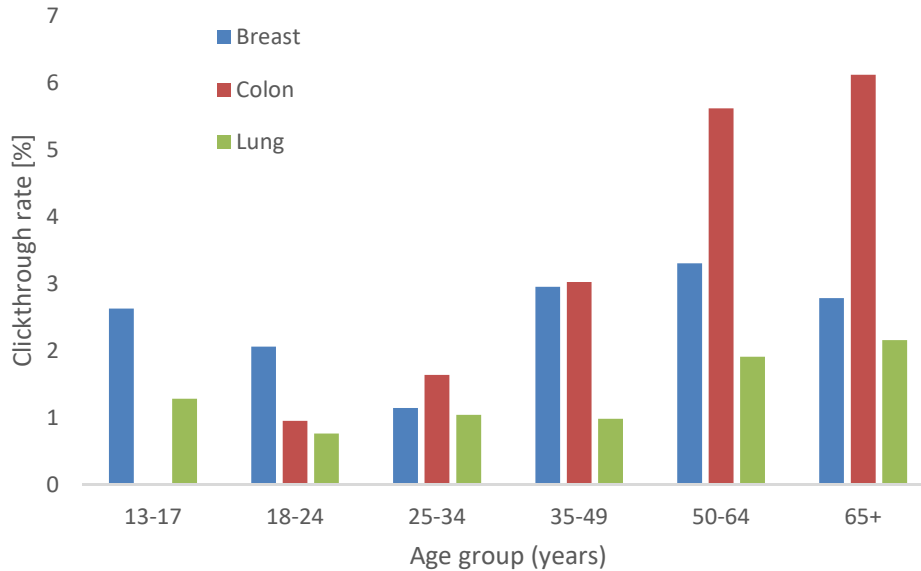}
  \caption{Clickthrough rates on ads, by cancer type and age group}
  \label{fig:ctr}
\end{figure}

Clickthrough rates on ads, by cancer type and age group, are shown in Figure \ref{fig:ctr}. As the figure shows, while the range of clickthrough rates are similar, older people tended to click more on ads, with the exception of breast cancer, which was also clicked by younger people. While the former is understood through the higher incidence of cancer in older ages, we attribute the latter to puberty, whose symptoms might be misinterpreted by some people as a serious disease. 

Throughout the experiment, 1285 questionnaires were started and 681 were completed (53\% completion rate). It took an average of 126 seconds to complete the questionnaires. After excluding people who did not consent to the participate in the study and people who did not have a query history of at least 14 days, a total of 308 people were analyzed. 

\begin{figure}
\centering
  \includegraphics[width=4.5in]{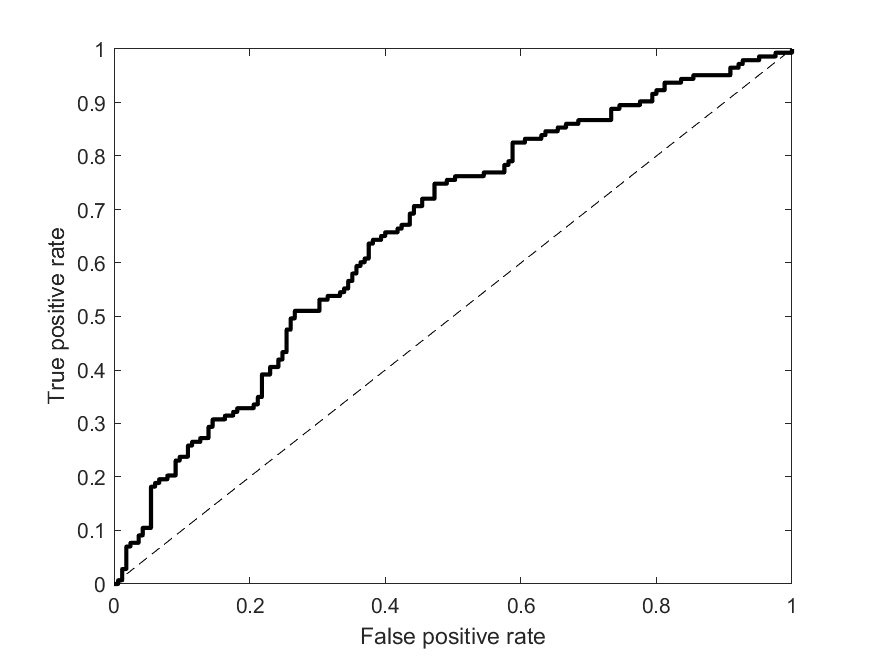}
  \caption{Receiver Operating Curve for detecting people with high likelihood of suspected cancer. The area under the curve is 0.66.}
  \label{fig:roc}
\end{figure}

Since the number of people who were screened for the different cancers were not equal, and some cancers had relatively few examples, we first attempted to predict the SCS for all cancers together. Figure \ref{fig:roc} shows the Receiver Operating Curve (ROC) \cite{duda2012} for the detection of all 3 cancers from the queries. As the figure shows, the Area Under the ROC curve (AUC) is 0.66, indicating that it is possible to identify those people who are very likely to have suspected cancer. 

Trained separately, AUC for the different cancers are 0.74 (colon), 0.56 (lung), and 0.50 (breast). Thus, colon cancer is the easiest one to identify, followed by lung cancer.

Attribute importance was calculated according to the average increase in prediction error if the values of that variable are excluded, divided by the standard deviation over the entire forest ensemble. The attributes most indicative of likely cancer were medically-related words ("`remedies"', "`colon"', "`pain"', and "`diet"') and words with no clear link to cancer ("`college"', "`joe"', "`north"', "`boots"', "`watch"' and "`movie"'). We hypothesize that these words typify specific demographics or behaviors. Importantly, these words did not discuss treatments or treatment centers, which would be expected if the people responding to the questionnaires were already post diagnosis.

\subsection{Study 2: Advertising systems can learn to identify people with suspected cancer}

\begin{figure}
\centering
  \includegraphics[width=4.5in]{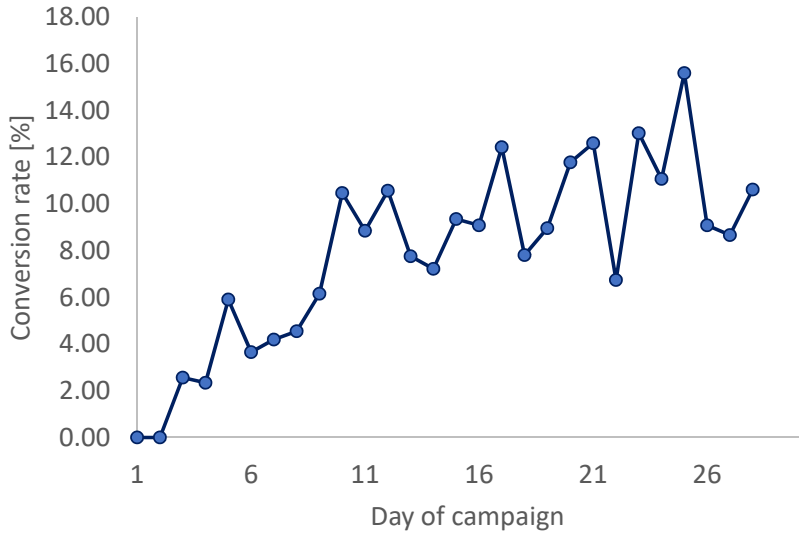}
  \caption{Conversion rate over time.}
  \label{fig:conversionrate}
\end{figure}

The ads were shown to 70,586 people during the period of May 16th to June 12th, 2018. During that time, 6,484 people clicked on these ads. Of those, 2917 people began the questionnaire and 1049 completed it (36\% completion rate). The conversion rate over time is shown in Figure \ref{fig:conversionrate}. As can be seen in the figure, the conversion rate rises from a negligible level to an average of 11\% (s.d. 3\%) in the last 10 days of the study. This means that the advertising system learned to identify people who would score high on the questionnaire, such that about one person in 9 who click on the ads are suspected to have cancer. The average conversion rates at the last 10 days of the study, for individual cancer types was 11\% for breast cancer, 9\% for colon cancer and 9\% for lung cancer. The number of people shown the ad each day was, on average, 2681 (s.d. 467), indicating that a relatively constant number of people saw the ads each day, but those people who saw the ads were increasingly more likely to have suspected cancer. Similarly, the rate of clicks on ads remained approximately constant at 10\% after the first 10 days of the study.

We modeled the data from 40 countries where ads were shown at least 150 times. A linear model of the click-through rate per country as a function of GDP per person (log transformed), Internet penetration and life expectancy reached an $R^2$ of 0.21, and found that Internet penetration was statistically significantly correlated with the click-through rate (slope: $0.001$, p-value: $0.04$), as was life expectancy (slope: $-0.002$, p-value: $0.02$). Thus, people living in countries with better internet access and worse health outcomes were more likely to use the ads to seek diagnosis. 

\section{Discussion}

The potential for using search engine queries to screen for different cancer types has been demonstrated, albeit through the use of biased populations who indicated their condition in queries or for people whose condition was inferred. Here we used a clinically verified questionnaire to identify people with suspected cancer, and correlate their past queries with the outcomes of the questionnaires. Our results suggest that the proposed method, which is economical for medical authorities to operate, can assist people with reduced access to the health system in pre-diagnosis of serious medical conditions. Moreover, this is done without causing undue stress to people without suspected cancer. Additionally, since people are seeking information to assist them in self-diagnosis, even in countries with developed health systems, our method could help alleviate some of the effort required in this endeavor. 

The outcomes of our first study indicate that queries are predictive of SCS, especially for those people for which the strength of the prediction is highest. Having demonstrated that queries are indicative of high SCS, our second study demonstrated that SCS can serve as a conversion signal to the ads system, enabling it to learn to identify people suspected of having these conditions based on their past queries, and inform them of possible risk through the combination of ads and questionnaires. 

One limitation of this study is that, although the questionnaires are designed to identify people with suspected cancer, our data does not contain diagnostic information. Future work will focus on a followup with people who completed their questionnaire until after diagnosis. Another limitation of our study could be that the advertising system, rather than advertising to representative populations, is focusing on a specific sub-population both because we advertised only to people whose queries indicated that they were worried that they are suffering from cancer or because specific populations respond better to our advertisements. We focused on people who inquired about cancer diagnosis and not, for example, on people who queried for relevant symptoms, as advertising for the latter could unnecessarily worry, and symptoms could be too unspecific. However, additional studies are needed to verify if advertisements can be shown to people making other queries indicative of cancer, even if they are not expressly querying for it, without causing undue stress and false indications.

%What does the impression rate mean, given the gender balance?

\pagebreak
\bibliographystyle{abbrv}

\end{document}